\documentclass[11pt]{article}
\usepackage[utf8]{inputenc}
\usepackage[authoryear]{natbib}
\usepackage[a4paper]{geometry}
\usepackage[title]{appendix}
\usepackage{xcolor}
\usepackage{setspace}
\newcommand{\bb}[1]{\boldsymbol{#1}}
\newcommand{\btheta}{\bb{\theta}}
\newcommand{\bomega}{\bb{\omega}}
\newcommand{\bSigma}{\bb{\Sigma}}
\newcommand{\bigO}{\mathcal{O}}
\newcommand{\bn}{\textbf{n}}
\newcommand{\bu}{\bb{\textbf{u}}}
\newcommand{\bs}{\bb{s}}

\newcommand{\I}{I_{\bn}(\bomega)}
\newcommand{\Gn}{\mathcal{G}_{\bn}}
\newcommand{\imag}{\text{i}}
\newcommand{\tll}{\ell_{\text{true}}(\btheta)}
\newcommand{\dwll}{\ell_{\text{dW}}(\btheta)}
\newcommand{\wmle}{\widehat{\btheta}_{\text{W}}}
\newcommand{\dwmle}{\widehat{\btheta}_{\text{dW}}}

\newcommand{\cll}{\ell_{c}(\btheta)}
\newcommand{\autocov}{c_{\btheta}(\bu)}

\newcommand{\eI}{\overline{I}_{\bn}(\bomega;\btheta)}
\newcommand{\fA}{f_{\btheta, \bb{\delta}}(\bb{\omega})}
\newcommand{\matern}{M{\'a}tern }

\newcommand{\nugget}{\sigma^2_{\varepsilon}}

\usepackage{enumerate}
\usepackage[shortlabels]{enumitem}

\usepackage[export]{adjustbox}

\usepackage{algorithm}
\usepackage{algpseudocode}
\algrenewcommand\algorithmicrequire{\textbf{Input:}}
\algrenewcommand\algorithmicensure{\textbf{Output:}}

\usepackage{amsmath}
\usepackage{amsfonts}
\usepackage{graphicx}
\usepackage{booktabs, multirow}

\usepackage{amsthm}
\theoremstyle{definition}
\newtheorem{simulation}{Simulation} 
\usepackage{url}

\title{Calibrated Bayesian inference for random fields on large irregular domains using the debiased spatial Whittle likelihood}
\author{Thomas Goodwin\thanks{School of Economics, University of New South Wales.} \and Arthur Guillaumin\thanks{School of Mathematical Sciences, Queen Mary University of London.} \and Matias Quiroz\thanks{School of Mathematical and Physical Sciences, University of Technology Sydney.} \and Mattias Villani\thanks{Department of Statistics, Stockholm University.} \and Robert Kohn$^*$}
\date{}
\begin{document}

\maketitle

\begin{abstract}
    Bayesian inference for stationary random fields is computationally demanding. Whittle-type likelihoods in the frequency domain based on the fast Fourier Transform (FFT) have several appealing features: i) low computational complexity of only $\bigO(n \log n)$, where $n$ is the number of spatial locations, ii) robustness to assumptions of the data-generating process, iii) ability to handle missing data and irregularly spaced domains, and iv) flexibility in modelling the covariance function via the spectral density directly in the spectral domain. It is well known, however, that the Whittle likelihood suffers from bias and low efficiency for spatial data. The debiased Whittle likelihood is a recently proposed alternative with better frequentist properties. We propose a methodology for Bayesian inference for stationary random fields using the debiased spatial Whittle likelihood, with an adjustment from the composite likelihood literature. The adjustment is shown to give a well-calibrated Bayesian posterior as measured by coverage properties of credible sets, without sacrificing the quasi-linear computation time. We apply the method to simulated data and two real datasets.
\end{abstract}

\section{Introduction} \label{sec: intro}

The collection and analysis of spatial data is crucial in many fields, such as geology \citep{cressie1989geostatistics, matheron1963principles}, climatology \citep{berliner2000long,hrafnkelsson2003hierarchical}, and epidemiology \citep{tolbert2000air,best2000combining}. Technological advancements make it possible to cheaply collect and store large amounts of spatial data. It is therefore important to develop models and estimation methods that can handle large-scale spatial data in a computationally feasible way.

Computing the likelihood for Gaussian random fields involves solving computationally demanding large systems of equations. Approximate methods like those in \citet{anitescu2017inversion} and \citet{stein2013stochastic} focus on estimating equation approaches to side-step expensive matrix computations via optimization and stochastic approximations, respectively. \citet{stroud2017bayesian} and \citet{guinness2017circulant} use circulant embedding techniques. \citet{siden2021spatial} use fast preconditioned conjugate gradient solvers. The class of stochastic partial differential equation (SPDE) models in \citet{lindgren2011explicit} approximates Gaussian random fields with the \matern covariance kernel by sparse and numerically efficient Gaussian Markov random fields \citep{rue2005gaussian}. Finally, so-called Vecchia approximations utilizing conditional independence (given a number of neighbors) of the field have been proposed to approximate the likelihood \citep{vecchia1988estimation, Katzfuss2021general}.

Another line of research uses the Whittle likelihood \citep{whittle1954stationary} to approximate the likelihood in the spectral domain \citep{gelfand2010handbook}. Whittle-type likelihoods have several appealing features: i) low computational complexity of only $\bigO(n \log n)$ from using the fast Fourier Transform (FFT), where $n$ is the number of spatial locations \citep{gelfand2010handbook}, ii) robustness to assumptions of the data-generating process, iii) easy handling of missing data and irregularly spaced domains \citep{fuentes2007approximate,matsuda2009fourier}, and iv) flexibility in modelling the covariance function directly in the spectral domain via the spectral density. Moreover, Whittle-type likelihoods open up for using efficient Markov chain Monte Carlo subsampling techniques \citep{quiroz2019speeding,quiroz2021block} in the frequency domain to further speed up posterior sampling algorithms both for univariate  \citep{salomone2020spectral} and multivariate \citep{villani2024spectral} data. The Whittle likelihood was initially developed for linear Gaussian models but has been extended to more general settings \citep{shao2007asymptotic}.

When used for time series data, the Whittle approximation is usually quite accurate with a small bias for the corresponding maximum likelihood estimator (MLE). Exceptions might occur due to a high dynamic range (the ratio between the maximum and minimum of the power spectrum), which can result in significant leakage, or due to not accounting for aliasing \citep{sykulski2019debiased}. In contrast, for two-dimensional (2D) data, for example, spatial data, the bias of the MLE based on the Whittle likelihood is substantial. \citet{guyon1982parameter} and \cite{kent1996spectral} prove that the bias incurred by the Whittle approximation is of the same order as the standard error in 2D, and that the bias is dominant for 3D spatial data; see also \cite{dahlhaus1987edge}.

\cite{guillaumin2022debiased} propose a debiased Whittle MLE by replacing the spectral density in the Whittle likelihood with the expected periodogram. They prove that their estimator has the same $\mathcal{O}(n^{-1/2})$ efficiency as the standard MLE in the time/space domain, regardless of the dimension of the lattice. However, because the debiased Whittle is a pseudo-likelihood, constructing a Bayesian posterior distribution based on the debiased Whittle likelihood is not guaranteed to be well-calibrated in the sense of  \cite{modrak2023simulation}, i.e.\ credible sets may not have the intended coverage.

Our paper makes the following contributions. We demonstrate that a Bayesian posterior distribution for the covariance kernel parameters based on the debiased Whittle likelihood can be badly calibrated \citep{modrak2023simulation} and will, therefore, not convey proper uncertainty quantification of the parameters. Using ideas from the composite likelihood literature in \cite{ribatet2012bayesian}, two alternative adjustments to the debiased Whittle posterior are presented. The adjusted debiased Whittle posteriors are empirically shown to be well-calibrated in a simulation study. Two real data applications are used to demonstrate the proposed method and to compare it against the standard Whittle posterior and the unadjusted debiased Whittle posterior.

The rest of the article is organized as follows. Section \ref{sec: debiased spatial whittle} reviews Whittle-type approximations to the exact Gaussian random fields likelihood and demonstrates that they result in poorly approximated posteriors. Section \ref{sec: posterior adjustments} reviews the concept of posterior calibration and proposes two methods for calibrating the debiased Whittle posterior, which are subsequently validated in a simulation study. Section \ref{sec: RF applications} demonstrates our method on two challenging large-scale datasets. Section \ref{sec: conclusion} concludes and outlines future research.

\section{Whittle-type likelihood approximations} \label{sec: debiased spatial whittle}
In this section, we review Whittle-type likelihood approximations that can be used to circumvent the
computational burden of the exact Gaussian likelihood. In particular, we discuss how the debiased spatial Whittle likelihood addresses several shortcomings of the standard Whittle likelihood. 

\subsection{The exact Gaussian likelihood}  

Let $X(\bs)\in\mathbb{R}$ be a finite variance, zero-mean random field indexed by the spatial location $\bs \in \mathbb{R}^d$ where $d\geq 1$ is a positive integer. Assume $X(\bs)$ is stationary --- but not necessarily isotropic --- and denote its covariance function by $c_{\btheta}(\bu)$, $\bu \in \mathbb{R}^d$, which is governed by some unknown parameters $\btheta \in \bb{\Theta}
\subset\mathbb{R}^p$ and is assumed to be square-summable. One of the most widely used covariance kernels is the isotropic \matern kernel,
\begin{equation} \label{eq: matern}
c_{\btheta}(\bu)=\sigma ^{2}{\frac {2^{1-\nu }}{\Gamma (\nu )}}{\Bigg (}{\sqrt {2\nu }}{\frac {||\bu||}{\rho }}{\Bigg )}^{\nu }B_{\nu }{\Bigg (}{\sqrt {2\nu }}{\frac {||\bu||}{\rho }}{\Bigg )},
\end{equation}
where $\Gamma(\cdot)$ and $B_{\nu}(\cdot)$ are, respectively, the Gamma function and the Bessel function of the second kind, and $||\cdot||$ is the Euclidean norm. The three parameters in $\btheta = (\rho,\sigma,\nu)^\top$ are the range $\rho$, amplitude $\sigma$ and smoothness $\nu$. 

Suppose that the observed random field $X_{\bs}$ is sampled on an orthogonal rectangular grid $\Gn$ with size $\bn = (n_1, \dots, n_d) \in (\mathbb{N}^+)^d$ and $|\bn|=\prod^d_{i=1} n_i$ denoting the total number of grid points. Furthermore, we denote the spacing between observations for each dimension as $\bb{\delta} = (\delta_1, \dots, \delta_d)$ with $|\bb{\delta}| = \prod^d_{i=1} \delta_i$. Throughout the simulations and real data examples, without loss of generality, we assume that the grid has a unit step size in all dimensions, i.e.  $\delta_1=\cdots = \delta_d=1$. Our methodology can also accommodate missing data and irregular sampling domains/boundaries, see below.

Let $p(\btheta)$ denote the prior of the kernel parameters. Exploring the posterior distribution $p(\btheta|X_{\bs}) \propto \mathcal{L}(\btheta) p(\btheta)$ requires computing the likelihood function $\mathcal{L}(\btheta)$. We use the term ``likelihood'' to refer to both the log-likelihood and likelihood, with the intended meaning clear from the context. The exact likelihood for Gaussian random fields is
\begin{equation} \label{eq: exact log-likelihood}
\tll =-\frac{|\bn|}{2}\log(2\pi)-\frac{1}{2}\log \text{det}\{\bSigma(\btheta)\}-\frac{1}{2}X^{\top}_{\bs}\bSigma^{-1}(\btheta)X_{\bs},
\end{equation}
where $\bSigma(\btheta)$ is the $|\bn| \times |\bn|$ covariance matrix corresponding to $c_{\btheta}(\bu)$ and $\text{det}\{\bSigma(\btheta)\}$ is the determinant of the covariance matrix. Evaluation of the likelihood is computationally challenging since $\text{det}\{\bSigma(\btheta)\}$
is $\mathcal{O}(|\bn|^{3})$ in general, or $\mathcal{O}(|\bn|^{5/2})$ in structured cases \citep{sowell1989decomposition, akaike1973block}. Furthermore, this likelihood is restrictive as it assumes the data-generating process is Gaussian, which may not be appropriate when modelling real data \citep{guilleminot2020modeling}. 

\subsection{The Whittle likelihood}

For any stationary random field, Bochner's theorem \citep{brockwell2009time} guarantees that there exists a spectral distribution function $F_{\btheta}(\bomega)$ such that
\begin{equation*}
    \autocov = \text{E}_{\btheta} \left[ X_{\bs} X_{\bs+\bu} \right] = \int_{\mathbb{R}^d} \exp(\imag \bomega \cdot \bu)d F_{\btheta}(\bomega), \quad \forall\bu\in\mathbb{R}^d,
\end{equation*}
where the expectation is for a fixed $\btheta$, $\bb{\omega} \in \mathbb{R}^d$ is a frequency vector, $\cdot$ is the dot product and we assume that $F_{\btheta}(\bomega)$ is absolutely continuous, such that it admits a spectral density function, $f_{\btheta}(\bomega) = \int_{\mathbb{R}^d} c_{\btheta}(\bu) \text{exp}(-\imag \bomega \cdot \bu)d \bu$. The \textit{aliased} spectral density $\fA$ of the sampled random field is,
\begin{equation} \label{aliased f}
    \fA = \sum_{\bu \in \mathbb{Z}^d} f_{\btheta}\left(\bb{\omega} + \frac{2\pi \bu}{\bb{\delta}}\right), \quad \quad \bb{\omega} \in \mathbb{R}^d.
\end{equation}
The periodogram of the sampled random field is given by
\begin{equation} \label{eq: RF I}
    \I = \frac{(2\pi)^{-d}|\bb{\delta}|}{\sum_{\bs\in \Gn}g_{\bs}^2}\left| \sum_{\bs\in \Gn}g_{\bs} X_{\bs}\text{exp}(-\imag\bomega \cdot \bs)\right|^2, \quad \quad \bomega \in \mathbb{R}^d,
\end{equation} 
where $g_{\bs}$, $\forall\bs \in \Gn$, is a masking grid which takes value $0$ if an observation is missing, and $1$ otherwise. Moreover, the masking grid $g_{\bs}$ also accommodates tapering, which can take values in the interval $[0, 1]$, with $0$ still indicating a missing value. Note that in the case of a fully observed grid with no tapering, the denominator in \eqref{eq: RF I} reduces to $|\bn|$. \cite{guinness2017circulant, stroud2017bayesian} use procedures that impute missing observations via circulant embedding, which may not be appropriate when the data violates the Gaussian assumption. Instead, the debiased Whittle handles missing observations and irregular sampling domains via the modulation values $g_{\bs}$~\citep{parzen_modulated_spectral_analysis}. The periodogram in \eqref{eq: RF I} is computed on a multidimensional grid of Fourier frequencies
\begin{equation*}
   \Omega_{\bn} = \prod^d_{j=1} 
    \left\{\frac{2\pi k}{\delta_j n_j} : k=0, \dots, n_j-1 \right\}.
\end{equation*}
The periodogram can be efficiently evaluated, regardless of missing observations or use of tapering, at a computation cost of $\mathcal{O}(|\bn|\text{log}|\bn|)$ operations via the FFT. The Whittle likelihood is a computationally efficient approximation to the Gaussian likelihood that relies on the following asymptotic result \citep{whittle1954stationary}
\begin{equation*} \label{eq: periodogram_asymp}
\I \overset{\mathrm{ind}}{\sim} \text{Exp}\left\{\fA\right\}, 
 \qquad \bomega \in \Omega_{\bn},
\end{equation*}
where $\text{Exp}(\lambda)$ is an exponential distribution parameterized by its mean $\lambda$. The Whittle likelihood is then
\begin{equation} \label{eq: wll}
\ell_{W}(\btheta) = -\frac{1}{2} \sum_{\bomega \in \Omega_{\bn}} \left\{ \text{log} \fA + \frac{\I}{\fA}\right\}.
\end{equation}

\cite{guyon1982parameter} and \cite{kent1996spectral} studied the accuracy of the Whittle likelihood approximation to the exact likelihood in \eqref{eq: exact log-likelihood} and prove that 
\begin{equation} \label{eq: ll compare}
    \left|\tll - \ell_{W}(\btheta)\right| = \mathcal{O}_P( |\bn|/n_1 )
\end{equation}
as $n_1 \rightarrow \infty$, where $n_1$ is the number of points on the smallest side of the lattice. Denote $\wmle$ as the maximum likelihood estimator of $\btheta$ using the likelihood in \eqref{eq: wll}.
From \eqref{eq: ll compare} it can be shown that bias of $\wmle$ is of the same order as the standard error when $d=2$, and for $d>2$ the bias is of larger order than the standard error. Hence, for $d \geq 2$, $\wmle$ is not an efficient estimator. For tapered data, \cite{dahlhaus1987edge} show that $\wmle$ is efficient for $d=2,3$. Despite the efficiency of $\wmle$ for tapered data, the rate of convergence is of the same order as the smallest side $n_1$. This is a drawback of the standard Whittle approximation as the bias of the MLE and its likelihood approximation is limited by the smallest side. Simulation studies suggest that the bias of $\wmle$ decreases slowly with respect to the grid size, see Figures 1 and 2 in \cite{guillaumin2022debiased}.

To assess the approximation of the likelihood function over $\Theta$ or around $\wmle$, we are interested in the relative order of error, i.e.\
\begin{equation} \label{eq: relative error}
    \left|\frac{\tll-\ell_{W}(\btheta)}{\tll}\right| = \mathcal{O}_{P}(1/n_1).
\end{equation}
The above equation illustrates the (probabilistic) convergence of the Whittle likelihood to the exact likelihood. From \eqref{eq: relative error}, we conclude that the relative error of the Whittle likelihood goes to zero (in probability) for square grids (which increase in all directions).

\begin{figure}
    \centering
    \includegraphics[width=13cm,height=13cm,keepaspectratio]{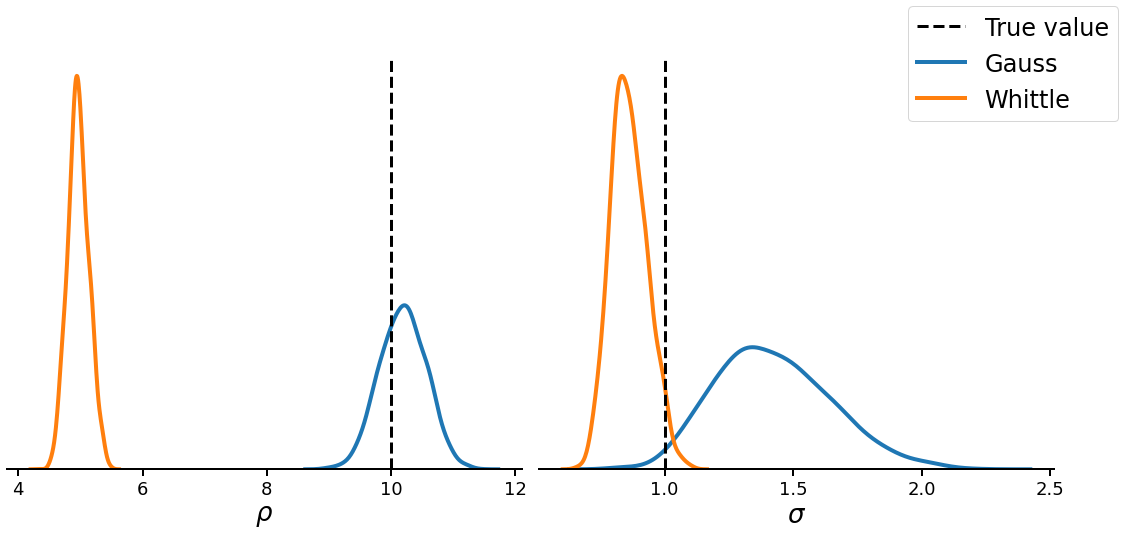}
    \caption{Kernel density estimates of the marginal posteriors based on different likelihoods (exact Gaussian, standard Whittle) for a simulated data example with a squared-exponential kernel with grid size $n=(64,64)$.}
    \label{fig: simulated example Gauss vs Whittle}
\end{figure}

To illustrate the poor posterior approximation based on the standard Whittle likelihood for spatial data, we simulate data on a square grid of $n=(64,64)$ from a Gaussian random field with the isotropic \matern kernel in \eqref{eq: matern}. We set the true values $\rho=10$ and $\sigma=1$, and fix $\nu =\infty$, which is not estimated. This corresponds to the squared-exponential kernel. Figure \ref{fig: simulated example Gauss vs Whittle} shows the marginal posterior density of $\rho$ and $\sigma$ from a non-informative prior for the exact Gaussian likelihood and the Whittle approximation.     
As seen in Figure \ref{fig: simulated example Gauss vs Whittle}, the Whittle posteriors are poor approximations of the posteriors from the exact Gaussian likelihood, particularly for the $\rho$ parameter. 

The following subsection presents the recently proposed debiased Whittle likelihood \citep{sykulski2019debiased,guillaumin2022debiased} and demonstrates that it requires modification to ensure well-calibrated Bayesian inference in terms of its posterior coverage properties.

\subsection{The debiased Whittle likelihood}

The debiased Whittle likelihood in \cite{sykulski2019debiased} for time series improves on the Whittle likelihood by reducing the bias. This is extended to domains of higher dimensions in \cite{guillaumin2022debiased}, e.g.\ for the analysis of spatial or spatiotemporal random processes. The debiased spatial Whittle likelihood is
\begin{equation} \label{eq: dwll}
\ell_{dW}(\btheta) = -\frac{1}{2} \sum_{\bomega \in \Omega_{\bn}} \left\{ \log \eI + \frac{\I}{\eI}\right\},
\end{equation}
where,
\begin{equation}\label{eq:expected_period_first_def}
    \eI = \text{E}_{\btheta} \left[ \I \right], \quad \quad \forall \bomega \in \Omega_{\bn},
\end{equation}
is the expected periodogram. Maximization of \eqref{eq: dwll} over $\Theta$ defines the debiased Whittle maximum likelihood estimator
\begin{equation*} \label{eq: mdwmle}
    \dwmle = \text{arg} \max_{\btheta \in \Theta} \dwll.
\end{equation*}

Equation \eqref{eq: dwll} shows that the debiased Whittle likelihood replaces the spectral density in the Whittle likelihood in \eqref{eq: wll} with the expected periodogram. The expected periodogram in \eqref{eq:expected_period_first_def} can be expressed as a convolution
\begin{equation} \label{eq: expected periodogram}
\eI = \left\{f_{\btheta, \boldsymbol{\delta}}* \mathcal{F}_{\bn} \right\}(\bomega) =  \int_{\mathcal{T}}  f_{\btheta, \bb{\delta}}(\bomega - \bomega^{\prime}) \mathcal{F}_{\bn, \bb{\delta}}(\bomega^{\prime})d\bomega^{\prime},
\end{equation}
where $\mathcal{T} = [0, 2\pi/\delta_1) \times \cdots \times [0, 2\pi/\delta_d)$, between the spectral density of the process and the multi-dimensional \textit{modified} F{\'e}jer kernel 
\begin{equation*} \label{eq: fejer kernel}
\mathcal{F}_{\bn}(\bomega) = \frac{(2\pi)^{-d}|\bb{\delta}|}{\sum_{\bs\in \Gn} g_{\bb{s}}^2} \left| \sum_{\bs\in \Gn}g_{\bs}\text{exp}(-\imag\bomega \cdot \bs)\right|^2, \qquad \bomega \in \mathbb{R}^d.
\end{equation*}
In the case of a fully observed domain with no tapering, this kernel becomes the multidimensional rectangular F{\'e}jer kernel, i.e.\ a separable product of one-dimensional F{\'e}jer kernels.

The debiased Whittle likelihood corresponds to the model
\begin{equation} \label{eq: parametric density}
\I \overset{\mathrm{ind}}{\sim} \text{Exp}\left\{\eI\right\}, 
 \qquad \bomega \in \Omega_{\bn}.
\end{equation}
Specifically, it ignores potential dependencies of the periodogram between distinct Fourier frequencies and, as such, can be considered misspecified and follows the framework of composite likelihoods \citep{varin2011overview}. \citet{guillaumin2022debiased} note that the estimator $\dwmle$ fits the methodology of estimating equations in \cite{heyde1997quasi}, since $\text{E}_{\btheta}[\nabla_{\btheta} \dwll] = \mathbf{0}$, where $\nabla_{\btheta}$ denotes the gradient with respect to $\btheta$, i.e.\ $\nabla_{\btheta} \dwll$ is the score function. \citet{guillaumin2022debiased} also prove that the debiased Whittle MLE has the same asymptotic efficiency as the time/space MLE, a result that holds for any dimension $d$ of the random field. 

The implementation of the expected periodogram in \eqref{eq: expected periodogram} can take advantage of the FFT and is therefore an $\mathcal{O}(|\bn|\log|\bn|)$ operation. This holds for any $d$, regardless of missing data or irregular domain patterns. The expected periodogram can directly account for finite sampling effects, such as aliasing and spectral leakage. The quantity $\eI$ is computed via the FFT of the discretely sampled theoretical covariance function combined with, in the case of fully observed grids, a multidimensional triangular kernel \citep{percival1993spectral}. Thus, the expected periodogram accounts for aliasing via discrete sampling of the covariance function and spectral leakage by inclusion of the triangular kernel. This is significant as the alternative of evaluating $\fA$ in \eqref{aliased f} as required in the standard Whittle likelihood estimation may be difficult \citep{sykulski2019debiased, guillaumin2022debiased}. For example, the aliased spectral density $\fA$ seldom has an analytical form and is usually approximated in practice (\citealt[Chapter 5]{gelfand2010handbook}) by truncation of the infinite sum in \eqref{aliased f} via \lq wrapping\rq \ contributions from $f_{\btheta}(\bb{\omega})$ for frequencies above the Nyquist frequency. This requires additional methods to account for aliasing, spectral leakage, missing data, and irregular sampling domains. All of these are automatically accounted for when computing $\eI$.

In the case of fully observed grids, \citet{guillaumin2022debiased} use Fejér's theorem~\citep{Korner_Fourier} to show that 
\begin{equation*} \label{eq: eI converges to spectral density}
\eI = \text{E}_{\btheta} \left[ \I \right]\xrightarrow[\bn\rightarrow\infty]{} f_{\btheta, \boldsymbol{\delta}}(\bomega), 
\end{equation*}
if $f_{\btheta,\boldsymbol{\delta}}$ is continuous,  where $\bn\rightarrow\infty$ denotes $n_i \rightarrow \infty$ for $i=1,\dots,d$. Hence, there is an asymptotic equivalence between the debiased Whittle and standard Whittle likelihoods. For more details on the expected periodogram and its computation, see  \cite{guillaumin2022debiased}.

\begin{figure}
    \centering
    \includegraphics[width=13cm,height=13cm,keepaspectratio]{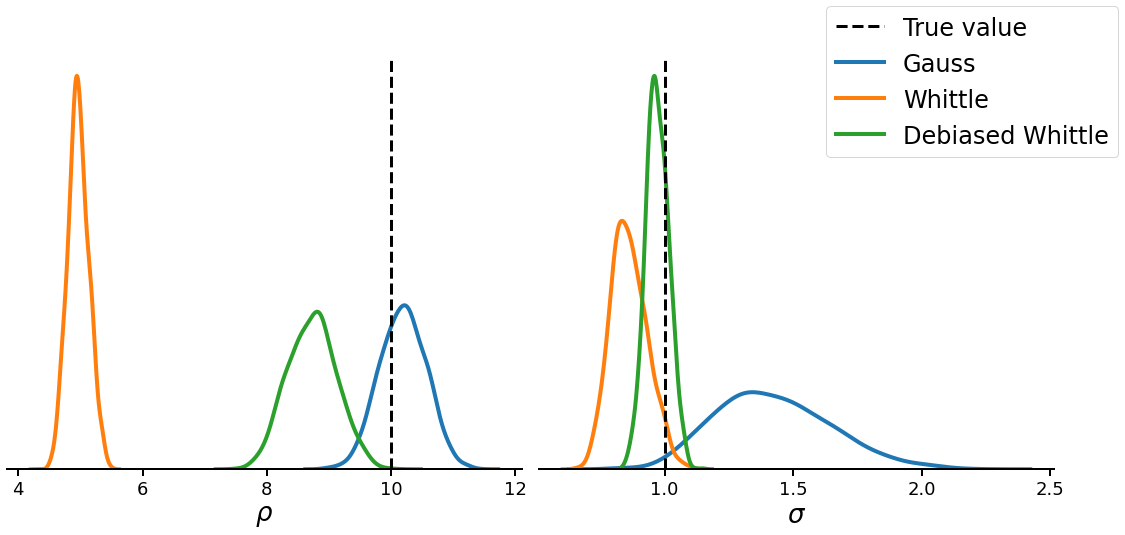}
    \caption{Kernel density estimates of the marginal posteriors based on different likelihoods (exact Gaussian, standard Whittle, debiased Whittle) for a simulated data example with a squared-exponential kernel with grid size $n=(64,64)$.}
    \label{fig: simulated example Gauss vs Whittle vs debiased Whittle}
\end{figure}

Figure \ref{fig: simulated example Gauss vs Whittle vs debiased Whittle} displays the marginal posteriors based on the three likelihoods (exact Gaussian, standard Whittle, debiased Whittle) with the same simulated data as in Figure \ref{fig: simulated example Gauss vs Whittle}. While the bias is reduced for the debiased Whittle posterior, the posterior uncertainty for $\sigma$ --- as is the case with the standard Whittle --- is small compared to the exact Gaussian posterior, a common phenomenon for composite likelihoods \citep{ribatet2012bayesian}. The next section shows how to adjust the posterior to obtain a better calibrated uncertainty quantification.

\section{Calibrating the debiased spatial Whittle posterior} \label{sec: posterior adjustments}
In this section, we investigate calibration methods to allow for valid Bayesian inference using the debiased spatial Whittle likelihood. We formally define valid posteriors, following
\cite{monahan1992proper}, and then propose curvature adjustments based on \cite{ribatet2012bayesian}. We present simulation studies that demonstrate the ability of the proposed methodology to leverage the debiased spatial Whittle likelihood for Bayesian inference.

\subsection{Well-calibrated posteriors}
We follow \cite{monahan1992proper} and define valid posteriors based on coverage properties of posterior sets. Assume the data are generated from the model $X \sim p(X_{\bs} | \btheta)$ and suppose $\mathcal{L}(\btheta ; X_{\bs})$ is the quasi-likelihood of interest, and we compute the quasi-posterior from
\begin{equation} \label{eq: bayes theorem}
    \widetilde{p}(\btheta | X_{\bs}) = \frac{\mathcal{L}(\btheta ; X_{\bs})p(\btheta)}{\widetilde{p}(X_{\bs})}, \quad \widetilde{p}(X_{\bs}) =  \int_{\Theta}\mathcal{L}(\btheta ; X_{\bs}) p(\btheta)d\btheta.
\end{equation}
Note that \eqref{eq: bayes theorem} corresponds to the Bayesian paradigm only when the likelihood comes from the data-generating process, i.e. $\mathcal{L}(\btheta ; X_{\bs})=p(X_s|\btheta)$ and then $$\widetilde{p}(\btheta | X_{\bs}) = p(\btheta | X_{\bs})\,\, \text{and} \,\,\widetilde{p}(X_{\bs}) = p(X_{\bs}) = \int_{\Theta}p(X_s|\btheta) p(\btheta)d\btheta.$$ 

Define $\widetilde{K}_{\alpha}(X_{\bs})$ as a posterior coverage set function of level $\alpha$ if, for every $X_{\bs}$,
\begin{equation} \label{eq:posterior_coverage_set_function} 
    \Pr\{\btheta \in \widetilde{K}_{\alpha}(X_{\bs})|X_{\bs}\} = \alpha, \quad (X_{\bs} \text{ conditioned upon}), 
\end{equation}
where the coverage set $\widetilde{K}_{\alpha}(X_{\bs})$ is determined from the quasi-posterior $\widetilde{p}(\btheta | X_{\bs})$ for a given $X_{\bs}$. Similarly, $K_{\alpha}(X_{\bs})$ denotes a posterior coverage set function when the set is determined from the posterior $p(\btheta | X_{\bs})$. A simple example of a posterior coverage set function is a $1-\alpha$ highest posterior density (HPD) interval for a one-dimensional parameter. The quasi-posterior $\widetilde{p}(\btheta | X_{\bs})$ is said to be valid by coverage if, for every $\alpha$,
\begin{equation} \label{eq:valid_coverage}
    \text{Pr}\{\btheta \in \widetilde{K}_{\alpha}(X_{\bs})\} = \alpha, \quad (X_{\bs} \text{ is random}),
\end{equation}
where the probability is now with respect to the joint distribution $p(X_{\bs}, \btheta) = p(X_{\bs}| \btheta) p(\btheta)$. Note the similarity to the classical statistical coverage, however, the sampling distribution of $X_{\bs}$ is averaged over the prior (instead of keeping the parameter at its true value). When the likelihood comes from the data-generating process ($\mathcal{L}(\btheta; X_{\bs}) = p(X_{\bs}|\btheta)$), the posterior is valid by coverage since 
\begin{align*}
    \text{Pr}\{\btheta \in K_{\alpha}(X_{\bs})\} & = \int_{X_{\bs}} \int_{\btheta \in K_{\alpha}(X_{\bs})} p(X_{\bs}, \btheta) d\btheta d X_{\bs} \\
    & = \int_{X_{\bs}} \int_{\btheta \in K_{\alpha}(X_{\bs})} p(X_{\bs}) p(\btheta | X_{\bs}) d\btheta d X_{\bs} \\
    & = \int_{X_{\bs}} p(X_{\bs}) \left( \int_{\btheta \in K_{\alpha}(X_{\bs})}   p(\btheta | X_{\bs})d\btheta \right) d X_{\bs} \\
    & = \int_{X_{\bs}} p(X_{\bs}) \alpha d X_{\bs}\\
    & = \alpha,
\end{align*}
using the definition of a posterior coverage set function in \eqref{eq:posterior_coverage_set_function} (with $K$ in place of $\widetilde{K}$). \cite{monahan1992proper} devise a method to test the validity of \eqref{eq:valid_coverage}, see Section \ref{sec: simulation study}.

We are interested in the case when $\mathcal{L}(\btheta; X_{\bs})$ differs from $p(X_{\bs}|\btheta)$, specifically when the debiased spatial Whittle likelihood in \eqref{eq: dwll} is used in place of the intractable Gaussian random field likelihood in \eqref{eq: exact log-likelihood}. Section \ref{sec: simulation study} demonstrates that the posterior based on the debiased spatial Whittle likelihood is not valid by coverage (and neither is the standard Whittle likelihood). We now turn to adjusting the approximate posterior based on ideas from the composite likelihood literature. The resulting adjustments are shown to dramatically improve its validity by coverage.

\subsection{Posterior adjustments}
 \cite{ribatet2012bayesian} propose an asymptotic curvature adjustment for composite likelihoods. The framework is closely related to maximum likelihood inference in misspecified models \citep{white1982maximum}. This curvature adjustment is inspired by the Bernstein Von-Mises theorem \citep{van2000asymptotic}, which loosely states that the posterior converges to the asymptotic sampling distribution of the MLE. See \cite{Kleijn2012Bernstein} for the Bernstein Von-Mises theorem in the case of misspecified models. Thus, the curvature adjustment corrects the variance of a composite posterior to equal (asymptotically) the variance of its corresponding maximum composite likelihood estimator (MCLE). In our case, \cite{guillaumin2022debiased} show that for finite grid sizes, the debiased spatial Whittle likelihood is a composite likelihood \citep{varin2011overview, bevilacqua2015comparing} and fits within the framework of estimating equations \citep{heyde1997quasi}. We briefly review how to perform these adjustments for composite likelihoods.

Let $\ell^{\mathrm{tot}}_c(\btheta)$ denote the composite likelihood of a sample with $n$ observations, which assuming independence decomposes as 
\begin{align}\label{eq:comp_likelihood_all}
    \ell^{\mathrm{tot}}_c(\btheta) & = \sum_{i=1}^{n} \ell_c(\btheta; y_i),
\end{align}
where $\ell_c(\btheta; y_i)$ denote the individual likelihood contributions for observations $y_i$, $i=1,\dots,n$. In the analogy to our case, the contributions correspond to the summands (multiplied by $-0.5$) in  \eqref{eq: wll} and \eqref{eq: dwll} with $y_i$ being the periodogram of the $i$th frequency and $n=|\bn|$. We use $\cll$ as a compact notation for $\ell_c(\btheta; y)$. Let $\widehat{\btheta}_c$ be the maximum composite likelihood estimator (MCLE) obtained by maximizing \eqref{eq:comp_likelihood_all}. Moreover, denoting the true parameter as $\btheta_0$ and 
the Hessian with respect to $\btheta$ as $\nabla^2_{\btheta}\cll$, let
\begin{equation*}
    \bb{H}(\btheta_0) = - \text{E}_{\btheta}\left[ \nabla^2_{\btheta}\cll \big|_{\btheta = \btheta_0} \right], \quad \bb{J}(\btheta_0) =\text{Var}_{\btheta}\left[ \nabla_{\btheta}\cll\big|_{\btheta = \btheta_0} \right],
\end{equation*}
denote the Fisher information matrix and the covariance of the score function, respectively. For proper likelihoods, $\bb{H}(\btheta_{0}) =\bb{J}(\btheta_{0})$ (under the usual regularity conditions) and the asymptotic distribution of the MLE  $\widehat{\btheta}$ is
\begin{equation*} \label{eq:MLE_asymp_dist}
    \sqrt{|\bn|}\bb{H}(\btheta_0)^{1/2}(\widehat{\btheta} - \btheta_0) \xrightarrow{d} \text{N}(\bb{0},\bb{I}).
\end{equation*}
For composite likelihoods on the other hand, $\bb{H}(\btheta_{0}) \neq \bb{J}(\btheta_{0})$, and the asymptotic distribution of the MCLE $\widehat{\btheta}_c$ is
\begin{equation} \label{eq: MCLE distribution}
    \sqrt{|\bn|}\bb{G}(\btheta_0)^{1/2}(\widehat{\btheta}_c - \btheta_0) \xrightarrow{d} \text{N}(\bb{0},\bb{I}).
\end{equation}
where $\bb{G}(\btheta)^{1/2}$ is the matrix square root of the so-called \lq sandwich\rq \ variance matrix \citep{varin2011overview}, defined as
\begin{equation*}  \label{eq: sandwich matrices}
 \bb{G}(\btheta) = \bb{H}(\btheta)\bb{J}^{-1}(\btheta)\bb{H}(\btheta).
\end{equation*}

For large enough $|\bn|$, it can be shown (Appendix A of \citealp{ribatet2012bayesian}) that the composite posterior $\pi_c(\btheta) \propto \exp(\ell^{\mathrm{tot}}_c(\btheta))p(\btheta)$ is approximately 
\begin{equation} \label{eq: comp post}
    \pi_c(\btheta) \sim \text{N}\left(\btheta_0, |\bn|^{-1}\bb{H}^{-1}(\btheta_0)\right).
\end{equation}
We see that there is a mismatch in the posterior covariance of \eqref{eq: comp post} to what we expect if the posterior distribution converged to the sampling distribution of the MCLE, i.e.\ $|\bn|^{-1}\bb{G}^{-1}(\btheta_0)$ \citep{Kleijn2012Bernstein}. The composite marginal posterior distributions of the parameters tend to have too small posterior variance compared to the posterior; see Figure 1 of \cite{ribatet2012bayesian}. The idea in \cite{ribatet2012bayesian} is to rescale the samples from the composite posterior to adjust for this as follows.

Define the asymptotically curvature-adjusted composite likelihood as
\begin{equation*} \label{eq: curv adj}
    \ell^{\mathrm{tot}}_\text{curv}(\btheta) = \ell^{\mathrm{tot}}_c(\btheta^*),  \qquad \btheta^* = \widehat{\btheta}_c + \bb{C}(\btheta - \widehat{\btheta}_c), 
\end{equation*}
where $\bb{C}$ is a positive semi-definite adjustment matrix such that
\begin{equation*} \label{eq: C^T H C}
\bb{C}^{\top}\bb{H}(\btheta_0)\bb{C} = \bb{H}(\btheta_0)\bb{J}(\btheta_0)^{-1}\bb{H}(\btheta_0).
\end{equation*}
\cite{ribatet2012bayesian} propose $\bb{C}=\bb{M}^{-1}\bb{M}_A$ where $\bb{M}^{\top}_A \bb{M}_A = \bb{G}(\btheta_0)$ and $\bb{M}^{\top}\bb{M}=\bb{H}(\btheta_0)$. The purpose of this adjustment is for $\ell^{\mathrm{tot}}_\text{curv}(\btheta)$ at $\widehat{\btheta}_c$ to match the curvature of the large-sample density of $\widehat{\btheta}_c$ (recall the Bernstein Von-Mises theorem for misspecified models). As a consequence, the curvature adjustment changes the location of any local maxima except the global maximum at $\widehat{\btheta}_c$, which may not be appropriate if the posterior is multi-modal. The resulting asymptotic distribution of the curvature adjusted  posterior $\pi_{\text{curv}}(\btheta)\propto \exp(\ell^{\mathrm{tot}}_{\text{curv}}(\btheta))p(\btheta)$ (see Appendix A of \cite{ribatet2012bayesian}) is
\begin{equation*}
     \pi_{\text{curv}}(\btheta) \sim \text{N}\left(\btheta_0, |\bn|^{-1}\bb{G}^{-1}(\btheta_0)\right).
\end{equation*}

As shown in \cite{guillaumin2022debiased}, the variance of the sampling distribution of the maximum spatial debiased Whittle likelihood estimator (MdWLE) $\dwmle$ is
\begin{equation} \label{eq: dbw sandwich}
    \text{Var}_{\btheta}\left[ \dwmle \right] \approx \bb{G}^{-1}(\btheta), 
\end{equation} 
which has the same sandwich structure as the covariance of the MCLE in \eqref{eq: MCLE distribution}. \cite{guillaumin2022debiased} give an analytical form of the asymptotic distribution of the MdWLE for Gaussian random fields when the observation domain $\bn$ grows to infinity in all directions; however, this form is seldom reached in practice. \cite{simons2013maximum} give a practical large-sample case where the asymptotic form has not been reached. In addition, empirical findings via simulations and, in the case of missing data, prevent the use of the exact form of the asymptotic variance for posterior adjustments. We therefore propose two methods that estimate the adjustment quantities via simulation.

\subsection{Two curvature adjustments}
This section explains the computation of the sandwich matrix in \eqref{eq: dbw sandwich} and proposes two ways of computing the corresponding adjustment matrix $\bb{C}$ specific to the debiased spatial Whittle likelihood. The computation of the analytic form of \eqref{eq: dbw sandwich} is intractable for larger grids for the reasons explained below, and thus, we resort to Monte Carlo simulation when computing the adjustment matrix $\bb{C}$. The computation of the adjustment matrices is performed once before the MCMC simulation at a one-time cost.

\cite{guillaumin2022debiased} give an analytic form $\bb{H}$ of the Fisher information matrix
\begin{equation*}
    \bb{H}(\btheta) = \frac{1}{2}\sum_{\bomega \in \Omega_{\bn_k}} {\overline{I}_{\bn_k}(\bomega;\btheta)}^{-2} \nabla_{\btheta} {\overline{I}_{\bn_k}(\bomega;\btheta)}
    \nabla_{\btheta}{\overline{I}_{\bn_k}(\bomega;\btheta)}^{\top},
\end{equation*}
where $\nabla_{\btheta}{\overline{I}_{\bn}(\bomega;\btheta)}$ is the gradient of the expected periodogram. This gradient can be computed efficiently via the FFT and using the same procedure as the expected periodogram, replacing the covariance function with its gradient; see Equation (38) in \cite{guillaumin2022debiased} for more details.

The $i,j$th element in the variance of the score vector, i.e. $\text{Var}_{\btheta}[\nabla_{\btheta} \dwll]$, is given as 
\begin{equation*} \label{eq: variance of score}
    \text{cov}\Bigg[ \frac{\partial\dwll}{\partial\theta_i}, \frac{\partial\dwll}{\partial\theta_j} \Bigg] = 
    | \bn|^{-2} \sum_{\bomega_1,\bomega_2 \in \Omega_{\bn}}
    \frac{\text{cov}\big[ I_{\bn}(\bomega_1), I_{\bn}(\bomega_2)\big]}{\overline{I}^2_{\bn}(\bomega_1; \btheta), \overline{I}^2_{\bn}(\bomega_2; \btheta)}
    \frac{\partial \overline{I}_{\bn}(\bomega_1; \btheta)}{\partial\theta_i} \frac{\partial \overline{I}_{\bn}(\bomega_2; \btheta)}{\partial\theta_j}.
\end{equation*}
This exact form involves a summation over $|\bn|^2$ 
terms, with each term itself requiring the evaluation
of the covariance of the periodogram at two Fourier
frequencies. Thus, a naive implementation would
require $\mathcal{O}(|\bn|^3)$ elementary operations in
the case of a Gaussian random field, and might prove even
more challenging without the assumption of Gaussianity.
Making use of the fast Fourier transform, the computational cost
may be brought down to $\mathcal{O}(|\bn|^2\log|\bn|)$
for a Gaussian random field, but this remains intractable even for moderate grid sizes. Our adjustments that we now turn to circumvent these computational issues by approximating the variance of the score.

Our first adjustment is based on the Monte Carlo estimate of $\text{Var}_{\btheta}[\nabla_{\btheta} \dwll]$,
\begin{equation} \label{eq: Jhat}
\bb{J}(\btheta) \approx {\bb{\widehat{J}}}(\btheta)= 
\frac{1}{M-1}
\sum^M_{i=1} 
\left(
\bb{g}^{(i)} - \overline{\bb{g}}
\right)
\left(\bb{g}^{(i)} - \overline{\bb{g}} \right)^{\top},
\end{equation}
where $\bb{g}^{(i)} = \nabla_{\btheta}\ell_{\text{dW}}(\dwmle | X^{(i)}_{\bs})$ and $\overline{\bb{g}}$ denotes the sample mean of the estimates. Once the MdWLE given the observed data is found, we simulate $M$ random fields from the specified model conditional on $\dwmle$. Then, $\bb{\widehat{J}}(\btheta)$ is obtained as the empirical covariance estimator based on $M$ Monte Carlo samples of
the gradient $\nabla_{\btheta}\ell_{\text{dW}}(\dwmle | X^{(i)}_{\bs})$. Note that simulation of Gaussian random fields can be performed efficiently in $\mathcal{O}(|\bn|\text{log}|\bn|)$ time via circulant embedding \citep{dietrich1997fast}. The adjustment is obtained as
\begin{align} 
    \bb{C}_1 &= \bb{M}^{-1}\widetilde{\bb{M}}_A, \label{eq: C1} \\
    \widetilde{\bb{M}}_A^{\top} \widetilde{\bb{M}}_A = \bb{H}(\btheta_0)&\bb{\widehat{J}}(\btheta_0)^{-1}\bb{H}(\btheta_0), \quad  \bb{M}^{\top}\bb{M} = \bb{H}(\btheta_0), \label{eq: est sandwich}
\end{align}
where $\widetilde{\bb{M}}^{\top}_A\widetilde{\bb{M}}_A$ and $\bb{M}^{\top}\bb{M}$ are singular value decompositions of their respective matrices. Note that $\widetilde{\bb{\bb{M}}}_A$ is an estimate of $\bb{M}_A$, however, since \eqref{eq: Jhat} converges in probability to $\bb{J}(\btheta)$ (weak law of large numbers) it follows that $\bb{H}(\btheta_0)\bb{\widehat{J}}(\btheta_0)^{-1}\bb{H}(\btheta_0)$ in \eqref{eq: est sandwich} converges in probability to $\bb{G}(\btheta_0)$ by the continuous mapping theorem (provided $\bb{J}(\btheta_0)$ is invertible). Thus $\widetilde{\bb{\bb{M}}}_A$ also converges in probability to $\bb{M}_A$.

The second adjustment has two key features. First, it directly estimates the sandwich matrix by a Monte Carlo estimate of $\text{Var}_{\btheta}[\dwmle]$, 
\begin{equation} \label{eq: MC var theta}
   \text{Var}_{\btheta}\left[\dwmle\right] \approx \widehat{\bb{G}^{-1}(\dwmle)} = 
\frac{1}{M-1}\sum^M_{i=1}\left(
\widetilde{\btheta}^{(i)}_{\text{dW}} - \overline{\widetilde{\btheta}_{\text{dW}}}
\right)
\left(
\widetilde{\btheta}^{(i)}_{\text{dW}} - \overline{\widetilde{\btheta}_{\text{dW}}}
\right)^{\top},
\end{equation}
where $\widetilde{\btheta}^{(i)}_{\text{dW}}$, for $i = 1,\dots, M$ are the MdWLE from $M$ datasets simulated at $\dwmle$. 
Simulating data $X^{(i)}_{\bs}$ from the likelihood at $\dwmle$ and finding the corresponding MdWLE for $M$ iterations gives the simulation approximation of the MdWLE distribution. Second, the observed Fisher information matrix, which we denote $\mathcal{H}(\btheta)$, of $\ell_{\text{dW}}(\btheta)$ is used in place of $\bb{H}(\btheta)$ when computing $\bb{M}$. The observed Fisher information is an unbiased estimator of the Fisher information, which matches the curvature of the unadjusted debiased Whittle likelihood, providing a tailored curvature adjustment for the specific dataset. Hence, the adjustment is
\begin{align}
    \bb{C}_2 &= \widehat{\bb{M}}^{-1}\widetilde{\bb{M}}_A, \label{eq: C2} \\
    \widetilde{\bb{M}}^{\top}_A \widetilde{\bb{M}}_A =  \widehat{\bb{G}(\btheta)}&, \quad  \widehat{\bb{M}}^{\top}\widehat{\bb{M}}=\mathcal{H}(\btheta), \label{eq: M_A M}
\end{align}
where $\widetilde{\bb{M}}^{\top}_A\widetilde{\bb{M}}_A$ and $\widehat{\bb{M}}^{\top}\widehat{\bb{M}}$ are singular value decompositions of their respective matrices.

We denote the curvature-adjusted debiased Whittle likelihoods as
\begin{equation} \label{eq: dbw curv adj}
    \ell_{\text{dW}}^{(i)}(\btheta) = \ell_{\text{dW}}(\btheta^*),  \qquad \btheta^* = \widehat{\btheta}_{\text{dW}} + \bb{C}_{i}(\btheta - \widehat{\btheta}_{\text{dW}}), \quad \text{for} \ i = 1,2.
\end{equation}
Combining $\ell_{\text{dW}}^{(i)}(\btheta)$ with a specified prior $p(\btheta)$ in a Metropolis-Hastings algorithm targets the desired posterior density $\pi_{\text{dW}}^{(i)}(\btheta)$ for $i = 1,2$. Algorithms \ref{alg: C1 adjustment algorithm} and \ref{alg: C2 adjustment algorithm} describe the computations necessary to obtain adjustments $\bb{C}_1$ and $\bb{C}_2$, respectively.

\subsection{Recommendations} \label{subsection: recommendations}
We now discuss some important considerations and recommendations for both curvature adjustments. The $\bb{C}_1$ matrix requires the evaluation of  $\nabla_{\btheta}\dwll$, which requires the gradient of the covariance function wrt the parameters. It is well known that the gradient of the \matern kernel in \eqref{eq: matern} wrt $\nu$ exists analytically but is difficult and computationally expensive. For this reason, we restrict the use of $\bb{C}_1$ for fixed $\nu$; however, it is possible to apply approaches such as \cite{geoga2023fitting} for the \matern kernel. 

Both adjustments employ Monte Carlo estimation. The variance of the Monte Carlo estimates is an important consideration when choosing $M$ to compute the adjustments. Thus, for cases when the variance is higher, a larger $M$ is required. Generally, small and moderate grid sizes require larger $M$, e.g.\ $M=1000$. 

The $\bb{C}_1$ adjustment performs a Monte Carlo estimate of $\text{Var}_{\btheta}\left[ \nabla_{\btheta}\dwll\right]$, whereas the $\bb{C}_2$ adjustment estimates $\text{Var}_{\btheta}\left[\dwmle\right]$. We have observed that the variance of $\text{Var}_{\btheta}\left[\dwmle\right]$ is large when the domain size is small relative to the value of $\rho$. In this case, we recommend using $\bb{C}_1$ since  $\text{Var}_{\btheta}\left[ \nabla_{\btheta}\dwll\right]$, and the variance thereof is less sensitive to $\rho$ compared to the domain size. However, for more difficult settings such as missing data and irregular domains, the estimate of $\text{Var}_{\btheta}\left[\dwmle\right]$ may be a better approximation of $\bb{G}^{-1}(\btheta)$ (or its inverse) compared to the $\bb{C}_1$ adjustment. Additionally, the computation of $\nabla_{\btheta}\dwll$ for $\bb{C}_1$ may not be known in closed form for missing data and/or irregular domains. Note that while the adjustment based on $\bb{C}_2$ may be more accurate, it is also computationally more burdensome, particularly for larger grids, due to the optimization of the likelihood for each $\dwmle^{(i)}$ in \eqref{eq: MC var theta} which can require multiple evaluations of $\dwll$.

Another important phenomenon is when the adjusted likelihood becomes flat. If elements of $\bb{C}$, particularly on the diagonal, are small, the vector $\bb{C}(\btheta - \widehat{\btheta}_{\text{dW}})$ in \eqref{eq: dbw curv adj}, becomes small. Thus for a proposed $\btheta$ that is far from $\widehat{\btheta}_{\text{dW}}$, the adjusted parameter $\btheta^*$ will still be close to $\widehat{\btheta}_{\text{dW}}$. As a consequence, the adjusted likelihood becomes flat around $\widehat{\btheta}_{\text{dW}}$, and the corresponding posterior will be dominated by the prior. To minimize the influence of the prior on the adjusted likelihood, a non-informative prior such as the penalised complexity (PC) prior is recommended, described in more detail in Section \ref{sim: PC prior}.

\begin{minipage}{0.46\textwidth}
\begin{algorithm}[H]
    \centering
    \caption{Adjustment $\bb{C}_1$.}\label{alg: C1 adjustment algorithm}
    \begin{algorithmic}[1]
    \Require{MdWLE $\dwmle$}
        \For{$i = 1, \dots, M$}{ \textbf{in parallel}}
        \State \small{Simulate $X^{(i)}_{\bs} \sim p(X_{\bs} \ | \ \dwmle)$}.
        \State \small{Compute $\nabla_{\btheta}\ell_{\text{dW}}(\dwmle | X^{(i)}_{\bs}$)}.
        \EndFor
        \State Compute $\widehat{J}(\dwmle)$ from \eqref{eq: Jhat}.\phantom{aaaaaa} \phantom{aaaaaa}  \phantom{aaaaaa} 
        \State Factor \vspace{-10pt} \begin{align*}
           \bb{M}^{\top}\bb{M} & =\bb{H}(\dwmle) \\
           \widetilde{\bb{M}}^{\top}_A \widetilde{\bb{M}}_A & = \bb{H}(\dwmle)\bb{\widehat{J}}(\dwmle)^{-1}\bb{H}(\dwmle).
        \end{align*}
        \vspace{-20pt} 
        \State \textbf{Return}   $\bb{C}_1 = \bb{M}^{-1}\widetilde{\bb{M}}_A$.\vspace{4pt}
    \end{algorithmic}
\end{algorithm}
\end{minipage}
\hfill
\begin{minipage}{0.46\textwidth}
\begin{algorithm}[H]
    \centering
    \caption{Adjustment $\bb{C}_2$}\label{alg: C2 adjustment algorithm}

    \begin{algorithmic}[1]
        \Require{MdWLE $\dwmle$.}
        
        \For{$i = 1, \dots, M$}{ \textbf{in parallel}}
        \State \small{Simulate $X^{(i)}_{\bs} \sim p(X_{\bs} \ | \ \dwmle)$}. 
        \State \small{$\widetilde{\btheta}_{\text{dW}}^{(i)} = \text{arg} \min_{\btheta \in \Theta} \ell_{\text{dW}}(\btheta | X^{(i)}_{\bs}).$} 
        \EndFor
        \State \small{Compute $\text{Var}_{\btheta}\left[\dwmle\right] \approx \widehat{\bb{G}^{-1}(\dwmle)}$ from \eqref{eq: MC var theta}}.
        \State Factor \vspace{-10pt} \begin{align*}
           \widehat{\bb{M}}^{\top}\widehat{\bb{M}} & = \mathcal{H}(\dwmle) \\
           \widetilde{\bb{M}}_A^{\top} \widetilde{\bb{M}}_A & = \widehat{\bb{G}(\dwmle)}.
        \end{align*}
        \vspace{-20pt} 
        \State \textbf{Return}   $\bb{C}_2 = \widehat{\bb{M}}^{-1}\widetilde{\bb{M}}_A$.
    \end{algorithmic}
\end{algorithm}
\end{minipage}
\bigskip

Table 1: Algorithms for the adjustments $\bb{C}_1$ (Algorithm 1) and $\bb{C}_2$ (Algorithm 2).

\subsection{Simulation study} \label{sec: simulation study}

 We use the computational approach of \cite{monahan1992proper} to validate the coverages of the curvature adjustments in Section \ref{sec: posterior adjustments}. This approach was later formalized into a software validation algorithm in \cite{cook2006validation}. First, simulate $i=1,\dots, K$ independent 
 samples from the prior, $\btheta^{(i)} \sim p(\btheta)$, and for each $\btheta^{(i)}$, generate a random field $X^{(i)}_{\bs} \sim p(X_{\bs}|\btheta^{(i)})$ and compute the integral
\begin{equation} \label{eq: posterior quantile}
    U^{(i)} = \int^{\btheta^{(i)}}_{-\infty} \widetilde p(\btheta|X^{(i)}_{\bs}) d\btheta, \quad \text{for } i=1,\dots, K,
\end{equation}
where $\widetilde p(\btheta|X^{(i)}_{\bs})$ is a quasi-posterior.
A Monte Carlo estimate $\widehat{U}^{(i)}$ of the above integral is performed with $M$ samples from the quasi-posterior via the random walk Metropolis-Hasting algorithm. \cite{cook2006validation} proves that if $\widetilde p(\btheta|X^{(i)}_{\bs}) =  p(\btheta|X^{(i)}_{\bs})$ then $\widehat{U}^{(i)} \overset{d}{\rightarrow} \mathrm{Unif}(0,1)$ as $M\rightarrow \infty$, for each $i=1,\dots, K$. Hence, a QQ-plot of the $\widehat{U}^{(i)}$ against the uniform distribution can be used to detect miscalibration of a quasi-posterior.

Since the two proposed curvature adjustments are valid asymptotically, we construct a simulation study to verify this for increasing grid sizes. We consider various priors and covariance kernels. The three Whittle likelihoods considered are:
\begin{enumerate}
    \item $\ell_{\text{dW}}(\btheta)$, the unadjusted debiased spatial Whittle likelihood.
    \item $\ell^{(1)}_{\text{dW}}(\btheta)$, the $\bb{C}_1$-adjusted debiased spatial Whittle likelihood.
    \item $\ell^{(2)}_{\text{dW}}(\btheta)$, the $\bb{C}_2$-adjusted debiased spatial Whittle likelihood.
\end{enumerate}
For all simulations, $M = 250$ datasets are generated from a two-dimensional Gaussian random field with a \matern covariance kernel. 

\begin{simulation} \label{simulation 1}
Consider a two-dimensional Gaussian random field from a \matern covariance kernel with $\nu \rightarrow \infty$, known as the squared-exponential covariance kernel, defined as
\[
c(\bu|\rho, \sigma) = \sigma^2 \text{exp}\left(-\frac{||\bu||^2}{2\rho^2}\right).
\]
The spectral density is $2\pi \rho \exp(-2 \pi^2 \rho ^ 2 ||\bomega||^2)$ \citep{rasmussen2006gaussian}. This produces overly smooth simulations due to high spectral mass at the lower frequencies, whereas the higher frequencies have a negligible mass. Estimation with this kernel is difficult since the periodogram contains significant correlations between Fourier frequencies. We use independent Gamma priors, 
$\rho \sim \text{Gamma}(\alpha = 60, \beta = 10 ), \
\sigma \sim \text{Gamma}(\alpha = 60, \beta = 50 ),$ with means $\text{E}[\rho] = 6$ and $\text{E}[\sigma] = 1.2$, respectively. We also use square grids with no missing values and grid sizes $|\bn| = (256^2, 512 ^ 2, 1024 ^ 2)$.

Figure \ref{fig: qq plot full grids} displays the QQ-plots of the standard uniform distribution against $\widehat{U}^{(i)}$ for $i=1,\dots, K$ for each grid size and adjustment. The unadjusted debiased Whittle posterior (top row) yields quantiles that are far from uniform for all three grid sizes. The S-shape of the QQ-plots suggests that the unadjusted posteriors are heavily over-concentrated and therefore rarely cover the true parameter $\btheta_0$. The two adjusted posteriors, $\bb{C}_1$ in the middle row and the $\bb{C}_2$ adjustment in the bottom row, have dramatically better coverage properties than the unadjusted case. For the two larger grid sizes, the adjusted posteriors give almost perfectly calibrated posteriors. 

\begin{figure}[htp]
    \centering
    \includegraphics[width=15cm,height=15cm,keepaspectratio]{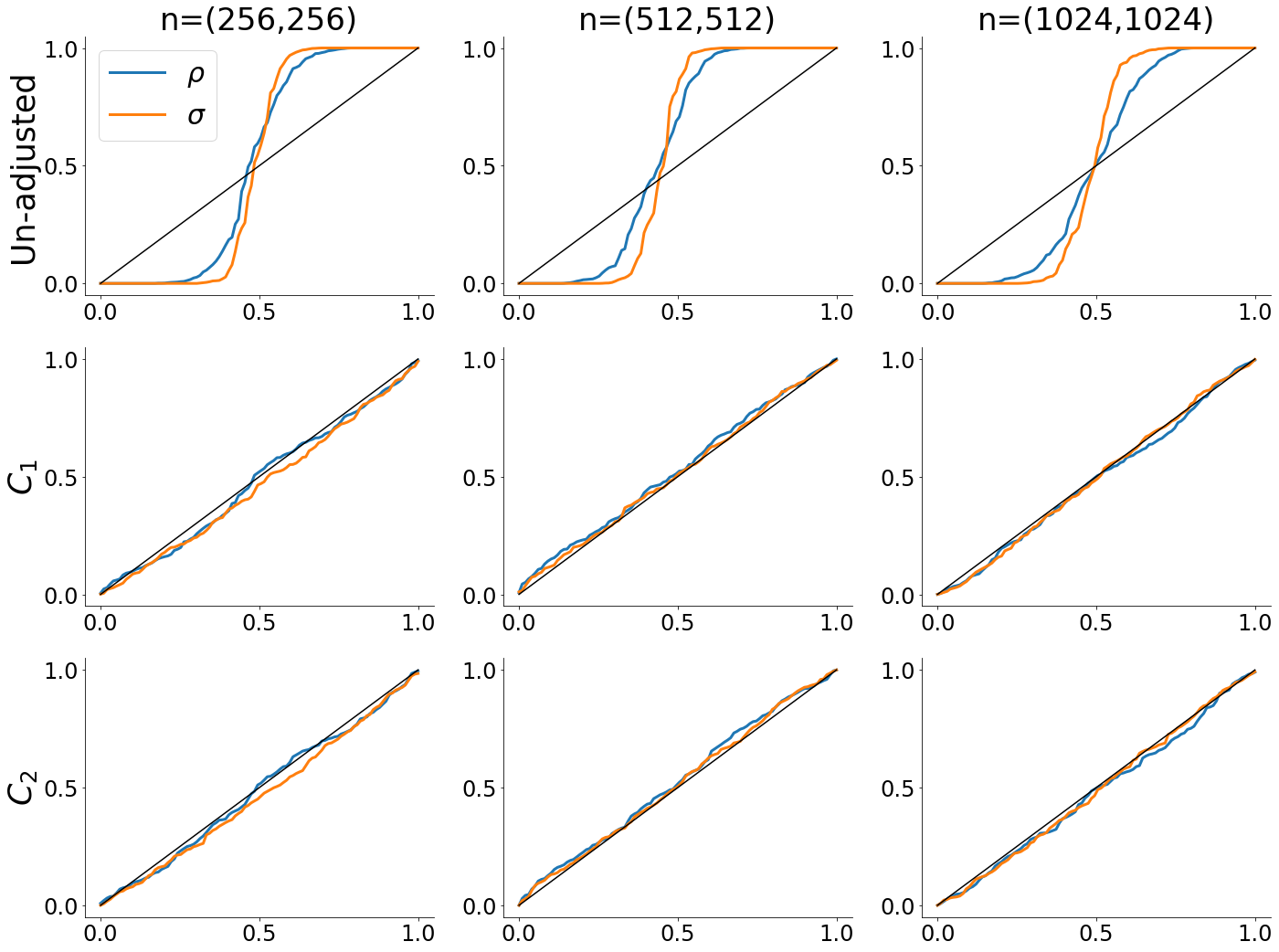}
    \caption{Simulation 1.~Standard uniform QQ-plots to quantify the coverage of quasi-posteriors for kernel hyperparameters of the Gaussian random field.}
    \label{fig: qq plot full grids}
\end{figure}
\end{simulation}

\begin{simulation}
We consider an irregular domain shape for this simulation. The domain shape is a grid of France. The grid is of size $\bn = (500, 500)$, and the 62\% of the grid points outside the border of France are treated as missing values. We use a squared-exponential covariance kernel with independent gamma priors, $\rho \sim \text{Gamma}(\alpha = 120, \beta = 20 )$, $\sigma \sim \text{Gamma}(\alpha = 50, \beta = 50 )$, with means $\text{E}[\rho] = 6$ and $\text{E}[\sigma] = 1$. We simulate $M=1000$ datasets to compute the $\bb{C}_1$ and $\bb{C}_2$ adjustments.

Figure \ref{fig: qq plots france} shows the QQ-plot of the posterior quantiles. As expected, the unadjusted debiased Whittle model fails to provide proper coverages for both parameters. The $\bb{C}_1$ and $\bb{C}_2$ adjustments are similar for both parameters, with $\bb{C}_2$ performing marginally better than $\bb{C}_1$. Both adjustments are indistinguishable from a standard uniform between $(0,0.5)$; however, the top half interval seems to have more concentration of mass compared to the bottom interval, with $\rho$ performing slightly worse than $\sigma$.  

\begin{figure}
    \centering
    \includegraphics[width=15cm,height=8cm,keepaspectratio]{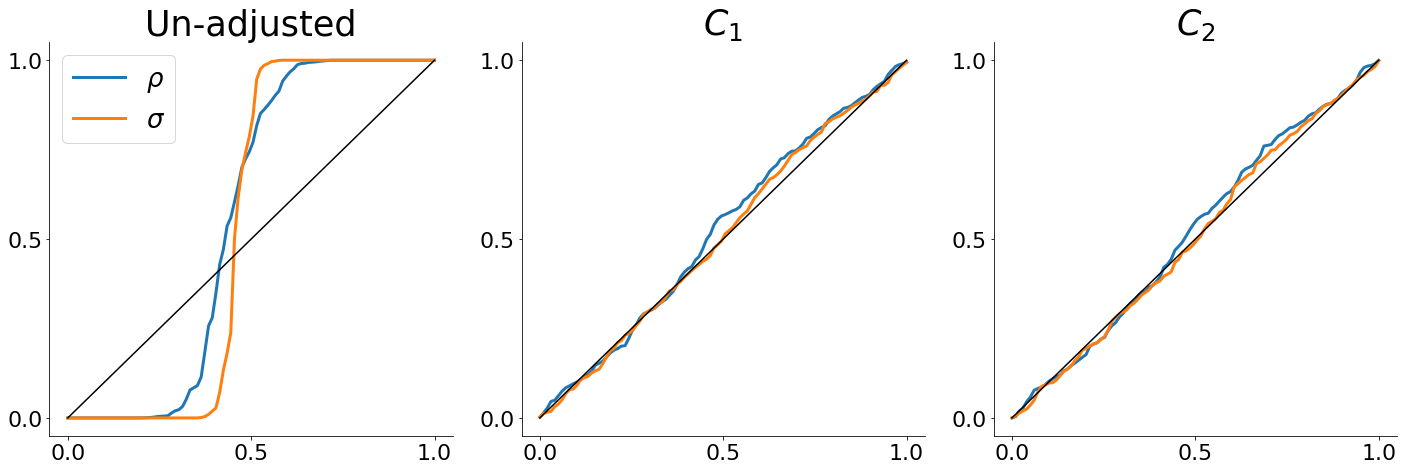}
    \caption{Simulation 2.~Standard uniform QQ-plots to quantify the coverage of quasi-posteriors for kernel hyperparameters of the Gaussian random field for an irregular domain shape of France.} 
    \label{fig: qq plots france}
\end{figure}
\end{simulation}

\begin{simulation} \label{sim: PC prior}
We consider the penalised complexity (PC) prior for random fields proposed in \cite{fuglstad2019constructing}. The PC prior is for \matern Gaussian random fields on the joint parameter space, range $\rho$ and amplitude $\sigma$, for a fixed smoothness $\nu=\infty$ (squared-exponential kernel). This prior can provide physically meaningful interpretations via its selection of hyperparameters. Furthermore, it is weakly informative and shrinks the range parameter towards infinity and the amplitude parameter towards zero. For two-dimensional random fields with fixed $\nu$, the density of the PC prior is given as
\begin{equation*}
    p(\rho, \sigma) = \lambda_1 \lambda_2 \rho^{-2} \exp(-\lambda_1 \rho^{-1} - \lambda_2 \sigma), \quad 
\end{equation*}
where $\rho, \sigma > 0,$ and
\begin{equation*}
    \lambda_1 = - \log(\alpha_1)p_0, \quad \text{and} \quad \lambda_2 = -\frac{\log(\alpha_2)}{\sigma_0},
\end{equation*}
with hyperparameters Pr$(\rho < \rho_0) = \alpha_1$ and Pr$(\sigma > \sigma_0) = \alpha_2$. To obtain samples from the prior, we take log transformations, $\log(\rho)$ and $\log(\sigma)$, and perform MCMC on the transformed space. For the simulation settings, we select hyperparameters $(\rho_0, \sigma_0, \alpha_1, \alpha_2) = (0.7, 1.0, 0.05, 0.05)$. Similar to Simulation \ref{simulation 1}, we consider square grids with no missing values with grid sizes $|\bn| = (256^2, 512 ^ 2, 1024 ^ 2)$.

Since the PC prior exhibits heavy tails for $\rho$, simulating data via circulant embedding and estimation is unfeasible when drawing large values of $\rho$ relative to the domain size. Thus, the corresponding quantiles in \eqref{eq: posterior quantile} are set to one for large proposed values of $\rho$. Additionally, the corresponding quantiles are set to zero for amplitudes that are extremely close to zero.

Figure \ref{fig: qq plot PC prior} displays the QQ-plots for the coverage with the PC prior. For $|\bn|=256^2$, the results for both $\bb{C}_1$ and $\bb{C}_2$ are close to uniform, whereas that of the unadjusted debiased Whittle posterior is not. All three posteriors are close to uniform for $|\bn|=512^2$, with both adjustments being almost indistinguishable from the standard uniform. In the case of $|\bn|=1024^2$, the $\bb{C}_1$ and $\bb{C}_2$ adjustments are very close to uniform, whereas the unadjusted is slightly worse for the $\rho$ parameter. Surprisingly, all posteriors for the $\sigma$ parameter are slightly worse for the largest grid size compared to the smaller grid sizes; note, however, that the posterior is naturally more concentrated for the largest grid size, so a slightly miscalibrated posterior variance can have a large effect.

\begin{figure}[htp]
    \centering
    \includegraphics[width=15cm,height=15cm,keepaspectratio]{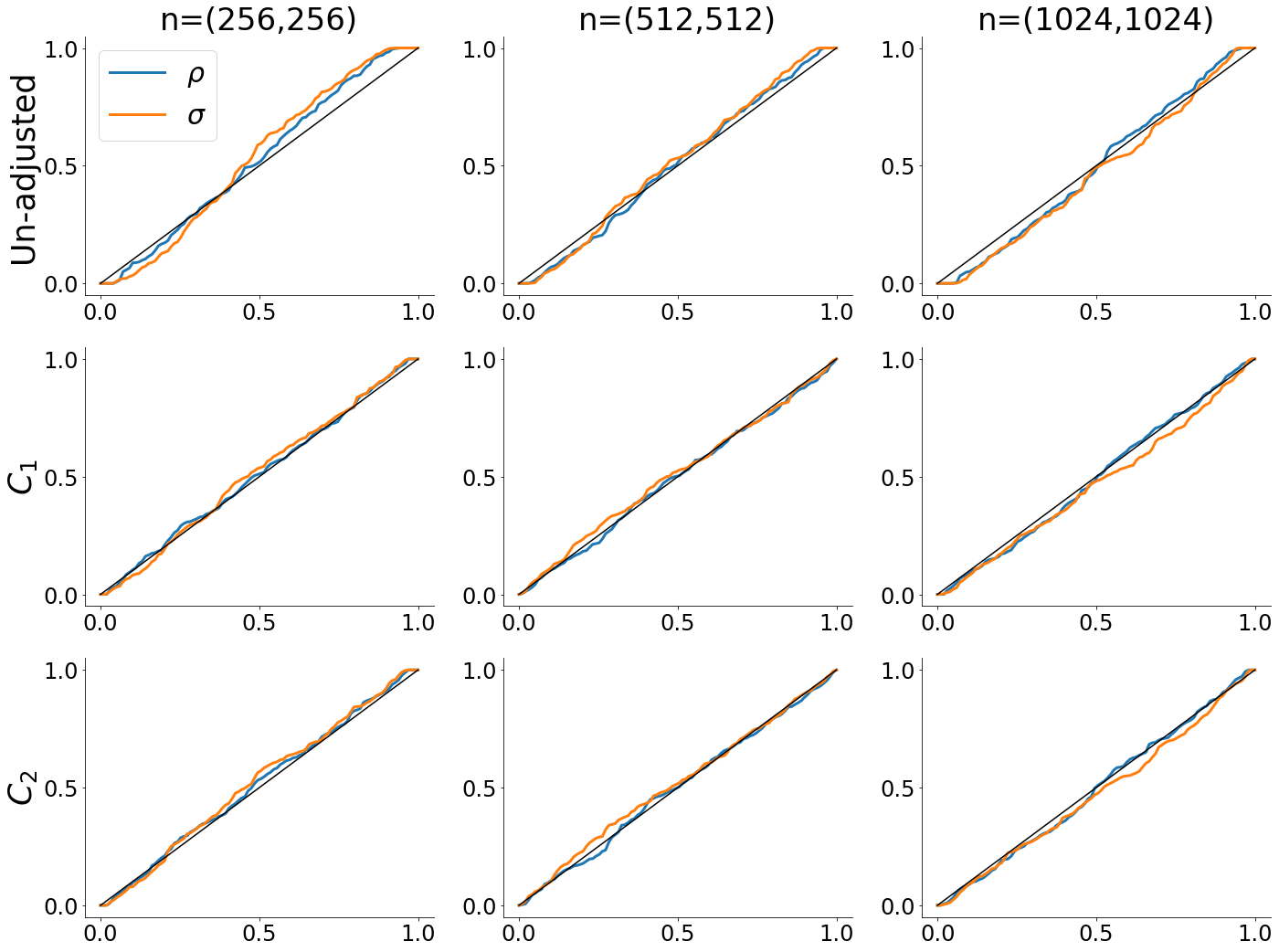}
    \caption{Simulation 3.~Standard uniform QQ-plots to quantify the coverage of quasi-posteriors for kernel hyperparameters of the Gaussian random field with the PC prior.}
    \label{fig: qq plot PC prior}
\end{figure}

\end{simulation}

\section{Applications} \label{sec: RF applications}

We illustrate our method on two popular datasets in the literature and compare our approach to the standard Whittle likelihood and the unadjusted debiased Whittle likelihood. The first example, sea surface temperature, is suited for the $\bb{C}_2$ adjustment since we assumed a \matern covariance kernel and also estimate the slope parameter $\nu$. As mentioned previously, the gradients of the \matern kernel wrt the parameters are difficult to compute, thus disqualifying $\bb{C}_1$. The second example, photosynthetically active radiation data, however, is suited for both adjustments. For a more detailed discussion of the suitability of each adjustment, refer to Section \ref{subsection: recommendations}.

\subsection{Sea surface temperature}

The first application is tropical rainfall measuring mission (TRMM) microwave imager (TMI) satellite data from the Pacific Ocean presented in Chapter 5 of \cite{gelfand2010handbook}. Sea surface temperature (SST) data are used for climate modelling and meteorology and are essential for evaluating climate change. They are also helpful for comparison with oceanic climate models as a diagnostic tool. Identifying spatial patterns of SST is a critical factor in the formation of hurricanes in the Pacific Ocean, which strike Central America. Furthermore, the transfer of water between the northern and southern equatorial currents is an important application of the analysis of the spatial structure of SST. Quantifying spatial variability and making informed predictions about SST is crucial for research on the world's ocean and the broader climate. The data are available at \url{https://images.remss.com/tmi/tmi_data_daily.html}.

The SST data, in degrees Celsius, is from March 1998 and has roughly a 25km $\times$ 25km spatial resolution defined by latitude and longitude. The data are observed on a rectangular grid of size 75 $\times$ 75. Due to the large number of observations, maximum likelihood estimation, let alone Bayesian inference via the Gaussian likelihood, is computationally expensive. Instead, we use the proposed Bayesian inference method using frequency domain methods.

\begin{figure}
    \centering
    \includegraphics[width=15cm,height=15cm,keepaspectratio]{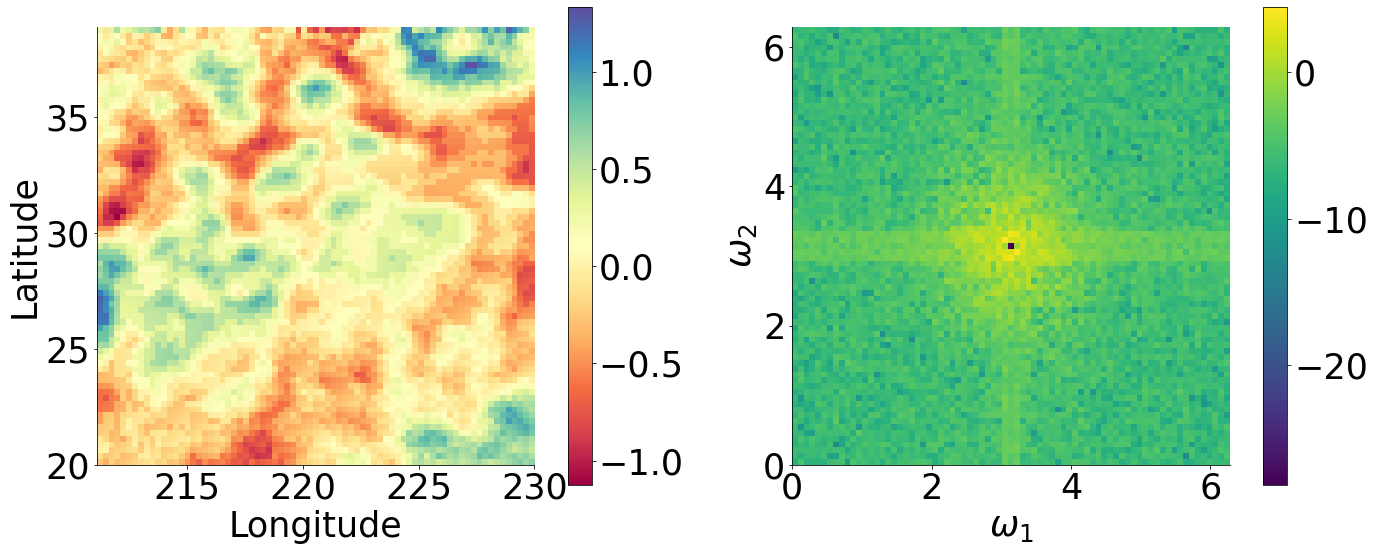}
    \caption{Sea surface temperatures over the Pacific Ocean. The left plots show the processed data after removing the trend. The right plot is the un-tapered log-periodogram of the data.}
    \label{fig: sst data}
\end{figure}

To satisfy the stationarity assumption, \cite{gelfand2010handbook} suggest removing a second-order polynomial mean trend 
\[
\beta_0 + \beta_1 u(\bs) + \beta_2  v(\bs) + \beta_3  u(\bs)^2 + \beta_4  v(\bs)^2 + \beta_5 u(\bs) v(\bs),
\]
where $u(\bs)$ and $v(\bs)$ are the longitude and latitude at each observation respectively. Figure \ref{fig: sst data} displays the stationary random field and its corresponding log-periodogram.

We also follow the suggestions from \cite{gelfand2010handbook} and use the \matern covariance kernel in \eqref{eq: matern} as an appropriate model. Initial optimisations are performed to obtain a sensible fixed value for the nugget parameter $\nugget=10^{-10}$. We perform Bayesian inference over the joint parameter space $\btheta = (\rho, \sigma, \nu)$. This is a challenging problem as it is well known that the smoothness parameter $\nu$ is difficult to estimate due to the lack of information in the likelihood \citep{de2022information}. 

We compare three posteriors: the unadjusted Debiased Whittle, the adjusted Debiased Whittle with $\bb{C}_2$ and the standard Whittle. Note that the $\bb{C}_1$ adjustment cannot be applied here as this requires the derivatives of the M{\'a}tern covariance wrt the parameters, which do not always exist. We simulate $M=500$ datasets to compute the adjustment $\bb{C}_1$ matrix. Independent Gamma priors are used for all three parameters,
\[
\rho \sim \text{Gamma}(\alpha = 5, \beta = 1.0 ), \
\sigma \sim \text{Gamma}(\alpha = 0.7, \beta = 1/0.7 ), \
\nu \sim \text{Gamma}(\alpha = 1.0, \beta = 2 ). 
\]

\begin{figure}[H]
    \centering
    \includegraphics[width=14cm,height=12cm,keepaspectratio]{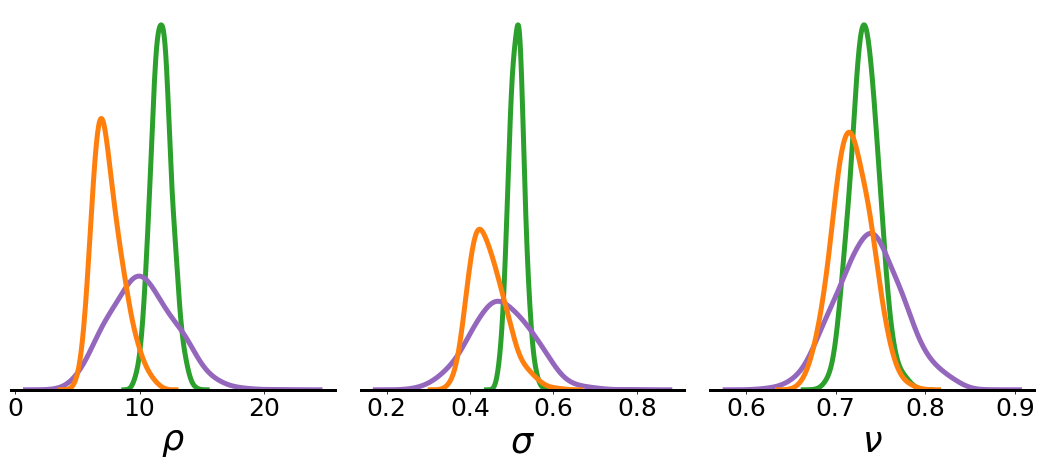}
    \caption{Sea surface temperatures: Kernel density estimates of the marginal posteriors of the kernel hyperparameters. The standard Whittle in orange, the unadjusted debiased Whittle in green, and the $\bb{C}_2$-adjusted debiased Whittle in purple.}
    \label{fig: sst posteriors}
\end{figure}

Figure \ref{fig: sst posteriors} plots the marginal posteriors from the different methods. The figure shows that the standard Whittle gives lower estimates of the $\rho$ and $\sigma$ than the debiased Whittle. The $\bb{C}_2$ adjustment in orange inflates the variance compared to the unadjusted debiased Whittle in blue. Overall, we conclude that the adjustment significantly reshapes the posterior. With support from the simulation studies in Section \ref{sec: simulation study}, this demonstrates the importance of the posterior adjustments in avoiding misleading inferences also on real data.

\begin{figure} 
    \centering
    \includegraphics[width=16.0cm,height=16.0cm,keepaspectratio]{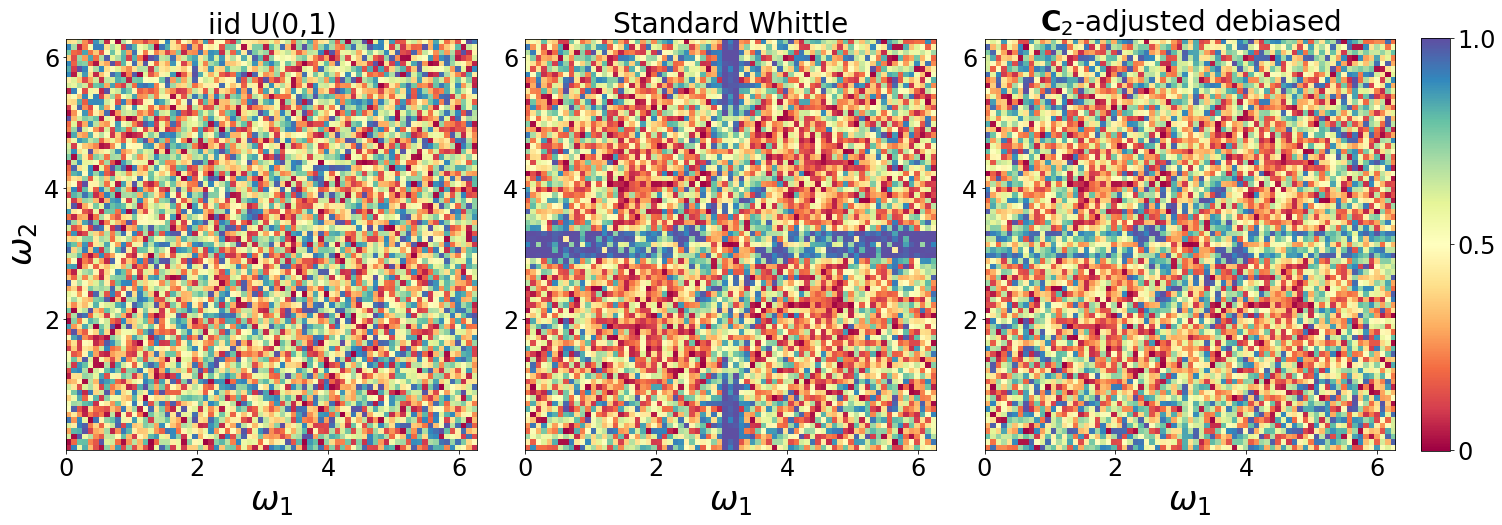}
    \caption{Residual spectrum $r_{\mathbf{n}}(\boldsymbol{\omega})$ in \eqref{eq: residual spectrum}. The left plot is the idealized iid standard uniform, the middle plot is the standard Whittle, and the right plot is $\bb{C}_2$ adjusted debiased Whittle. The estimated spectra are based on the posterior mean.}
    \label{fig: sst residuals}
\end{figure}

To check the adequacy of the model, we use the parametric model of \eqref{eq: parametric density}, and that if $X\sim \text{Exp}(\mu) \text{ then } U = 1-\exp(-X/\mu) \sim \mathrm{Unif}(0,1)$, to define the frequency domain residual spectrum as 
\begin{equation*} \label{eq: residual spectrum}   
r_{\mathbf{n}}(\boldsymbol{\omega}) = 1 - \exp\left( - \I \ / \ \eI\right) \overset{\mathrm{i.i.d.}}{\sim} \mathrm{Unif}(0,1), 
 \qquad \bomega \in \Omega_{\bn},
\end{equation*}
for the $\bb{C}_2$-adjusted debiased Whittle likelihood. Similarly, the residuals of the standard Whittle are obtained by replacing the expected periodogram with the spectral density. Ideally, given the correct model, the residuals have a standard uniform distribution throughout the entire observable domain. Figure \ref{fig: sst residuals} plots the standard Whittle residuals in the middle graph and the $\bb{C}_2$-adjusted debiased Whittle on the right using their associated posterior means; a simulation of the ideal case with a perfectly fitting model is shown on the leftmost graph as a reference. The residual spectrum in the middle graph from the standard Whittle model have very large and clearly visible side lobes, suggesting a poorly fitting model. The residual spectrum from the $\bb{C}_2$-adjusted debiased Whittle in the right hand graph has a little excess side lobes in the horizontal direction, but is overall quite close to the ideal iid $\mathrm{Unif}(0,1)$ case.

\subsection{Photosynthetically active radiation data}

We consider the NASA satellite data analysed in \cite{guinness2017circulant}. Photosynthetically active radiation (PAR) data contain the spectral range (in the interval 400-700 nanometers) of solar radiation used in photosynthesis. PAR is an integral part of primary producers such as phytoplankton, seagrass, plant growth, and species interaction. A deficiency of PAR can result in reduced growth or loss of corals, seagrass, and other photosynthetic organisms. Thus, PAR measurements give crucial insight into the quality and quantity of biological ecosystems. This dataset is from December 1, 2013, and can be found at https://oceandata.sci.gsfc.nasa.gov/l3/.


\begin{figure} 
    \centering
    \includegraphics[width=10.0cm,height=10.0cm,keepaspectratio]{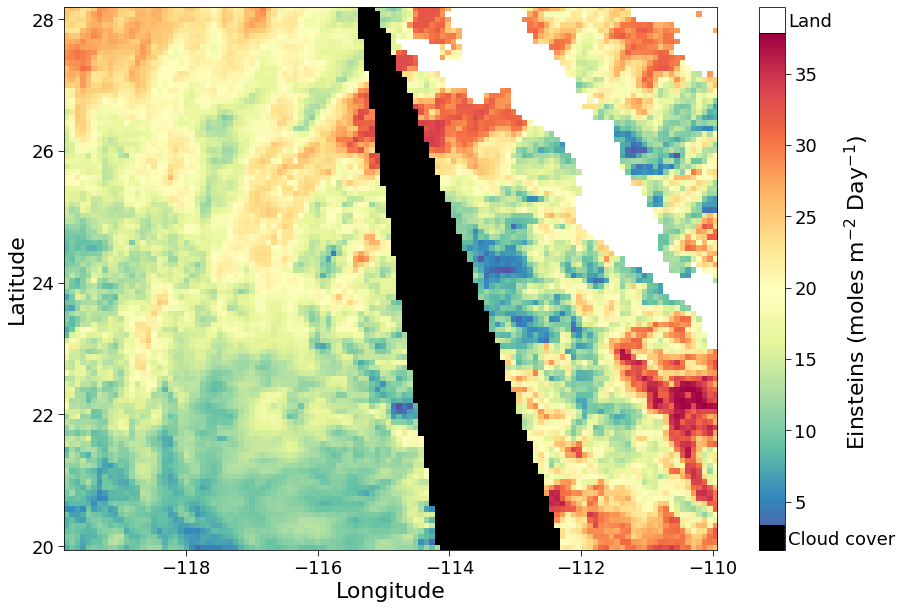}
    \caption{Photosyntehtically active radiation (PAR) data over the coast of Mexico in Baja California. The Black pixels represent missing data due to cloud cover, whereas the white pixels are missing data due to land.}
    \label{fig: radiance data}
\end{figure}

The data are located over the coast of Mexico, in Baja, California, with an evenly spaced grid size of 120 longitude values and 100 latitude values with an approximate resolution of 1/12\textdegree \ for both longitude and latitude. The data have a total number of lattice locations of $12000$. However, there are a total of 2270 missing values in the dataset for two reasons. First, PAR is measured over water, and thus, there are no readings over the land area. Second, the data are missing due to cloud cover. Our aim is to simulate the posterior of the parameters of the covariance kernel by applying the methods we developed and give some meaningful predictions of the missing observations due to cloud cover. The data are plotted in Figure \ref{fig: radiance data}. The large black triangular region contains missing values due to cloud cover, whereas the white values are missing due to land.

We follow \cite{guinness2017circulant} and subtract the empirical mean of the data since there are no apparent trends in the data. Furthermore, we use an exponential covariance kernel with unknown parameters $\btheta = (\rho, \sigma)$. As in the previous example, initial optimizations are performed to obtain a sensible fixed value for the nugget parameter $\nugget = 0.001$. We compare four methods: the standard Whittle likelihood, the un-adjusted debiased spatial Whittle likelihood, the $\bb{C}_1$-adjusted debiased spatial Whittle likelihood and the $\bb{C}_2$-adjusted debiased spatial Whittle likelihood. The parameters of interest $\btheta$ were transformed to the log-scale, and the PC prior was used for all methods with hyperparameter setting $(\rho_0, \sigma_0, \alpha_1, \alpha_2) = (0.7, 1.0, 0.05, 0.05)$, which is the same as in the simulation study.

\begin{figure} 
    \centering
    \includegraphics[width=12.0cm,height=12.0cm,keepaspectratio]{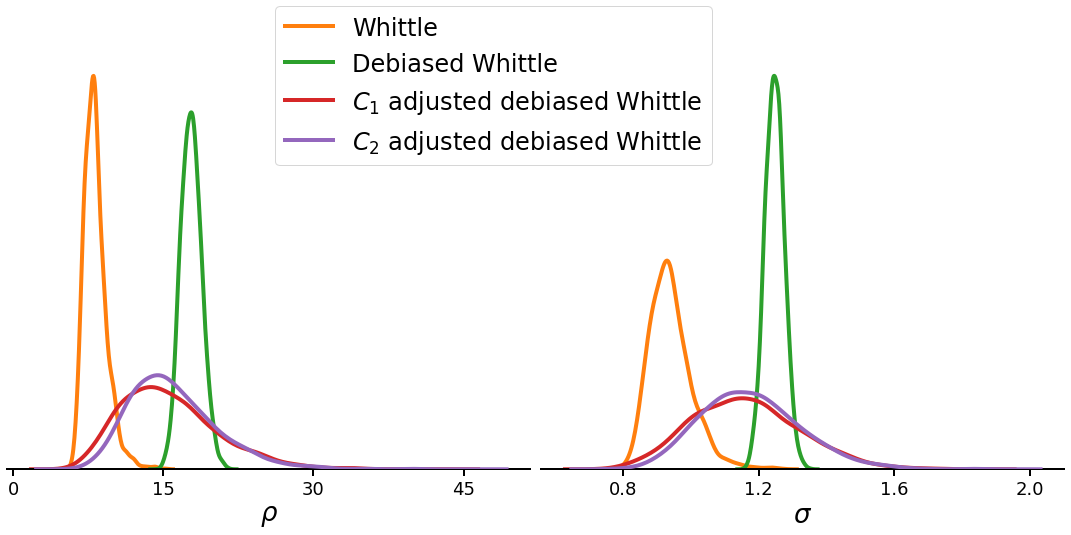}
    \caption{Kernel density estimates of the marginal posteriors for the photosynthetically active radiation data.}
    \label{fig: radiance posteriors}
\end{figure}

The kernel density estimates of the marginal posteriors are plotted in Figure \ref{fig: radiance posteriors}.  The standard Whittle posteriors have lower estimates for $\rho$ and $\sigma$ on average than its debiased Whittle counterparts. Furthermore, while the debiased Whittle posteriors correct for the bias of the Whittle likelihood, its posteriors are most likely too concentrated, judging from the simulation results in Section \ref{sec: simulation study}. The $\bb{C}_1$ and $\bb{C}_2$ adjusted debiased Whittle posteriors in red and purple, respectively, produce very similar marginal posteriors and have larger variances than the unadjusted debiased Whittle.

\begin{figure} 
    \centering
    \includegraphics[width=17.0cm,height=12.0cm,keepaspectratio]{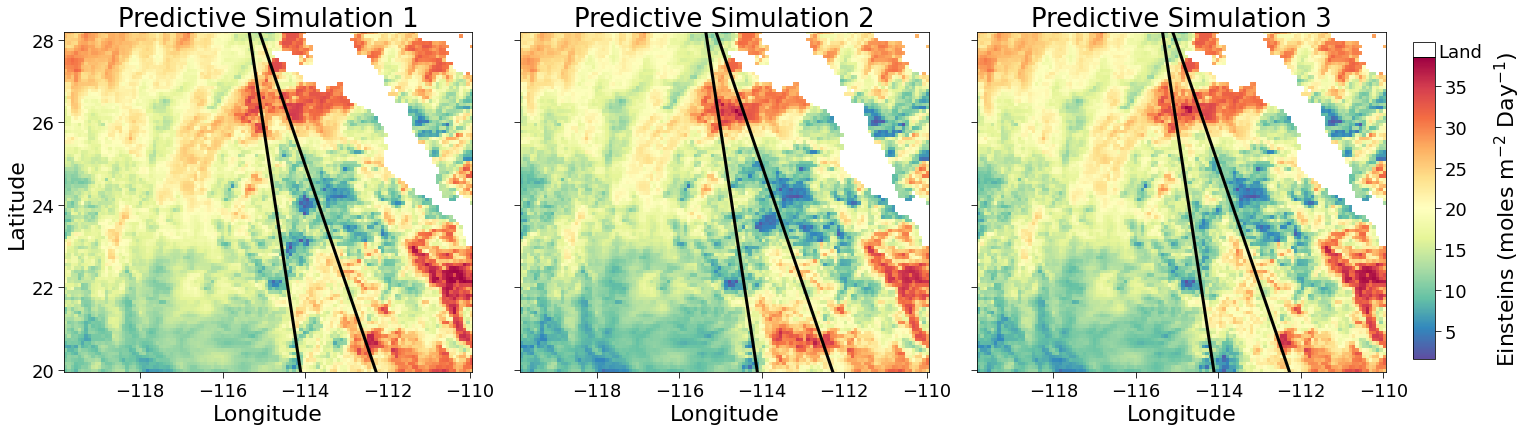}
    \caption{Three draws from the posterior predictive density with an exponential covariance model based on parameters from the MCMC chain at the $3000$th (left), $6000$th (middle), and $9000$th (right) iteration.}
    \label{fig: radiance predictive simulation}
\end{figure}

Interpolation over missing values of the PAR data provides insight into the uncertainty over the missing regions while still accounting for the covariance properties of the data. Similiar to \cite{guinness2017circulant}, Figure \ref{fig: radiance predictive simulation} shows three draws from the posterior predictive density, which it the conditional normal density \[p(\widetilde{X}(\bs_{\text{cc}}) | X(\bs), \btheta^{(i)}),\] where $i=3000, 6000, 9000$ are the $i$th iterate of the MCMC chain based on the $\bb{C}_1$-adjusted debiased Whittle, and $\bs_{\text{cc}}$ are the spatial locations of the missing data (triangular region indicated by the black lines) due to cloudcover. 

\section{Conclusion and future research} \label{sec: conclusion}

We propose a method to carry out Bayesian inference for covariance-stationary random fields using the computationally efficient debiased spatial Whittle likelihood recently proposed in the literature. In the classical paradigm, the debiased Whittle likelihood has many benefits discussed in \cite{guillaumin2022debiased} as opposed to other estimation methods (e.g.\ \citealp{whittle1954stationary}, \citealp{gelfand2010handbook}). However, because it is not a proper likelihood, its applicability to Bayesian inference needs careful treatment. Leveraging the fact that the debiased spatial Whittle likelihood falls within the framework of a pseudo-likelihood in finite samples, we use ideas from \cite{ribatet2012bayesian} to construct posterior curvature adjustments that asymptotically satisfy the Bernstein Von-Mises theorem for the subsequent adjusted debiased Whittle likelihood. These adjustments are designed to achieve appropriate coverage properties of the Bayesian posterior. The quantities needed to compute the posterior adjustments are not computationally tractable for the debiased spatial Whittle likelihood and we propose two methods to circumvent this. The first method works by computing the variance of the score function and is more robust to processes with large length scales. However, this adjustment relies on the derivative of the covariance function, which may not be analytically available for general \matern class covariance functions. The second method estimates the adjustment via a simulation approximation of the sampling distribution of the MdWLE and the observed Fisher information, which provides a tailored adjustment useful in problems with smaller grid sizes. We demonstrate that both adjustments dramatically improve the coverages of the approximate posterior in simulated examples.

We see at least five directions for future research. First, one particular strength of our method is that it does not require the data to be Gaussian, and can therefore be extended to non-Gaussian spatial models. Second, for ultra-large spatial data, spectral subsampling MCMC \citep{ salomone2020spectral, villani2024spectral} is an attractive alternative if the log-likelihood function can be computed independently for each data point. As it stands, this is not possible/inefficient for the debiased Whittle likelihood since the expected periodogram is computed on the whole grid of Fourier frequencies via the FFT, or in the frequency domain by a convolution that requires the whole spectrum, so some innovation is needed to make this practical. An alternative is to use some other type of bias correction of the standard Whittle likelihood in a subsampling context, combined with the coverage adjustments used in this paper. Third, our proposed approach handles missing data and irregular domains, but more work is needed to extend it to irregularly spaced data \citep{matsuda2009fourier}. Furthermore, although not explored here, our method can also be applied to time series models, i.e., when $d=1$ \citep{sykulski2019debiased}, and is thus a useful alternative to the standard Whittle estimator for time series data with a small number of observations. Finally, other methods of calibration can be explored. For example, the recent approach in \cite{frazier2023calibrated}.

\clearpage 

\bibliographystyle{apalike}
\bibliography{ref}

\end{document}